\documentclass[aps,prd,amssymb,eqsecnum,showpacs] {revtex4}
\usepackage{graphicx}
\usepackage{psfrag}

\begin{document}

\title{Binary Black Hole Waveform Extraction at Null Infinity}

\author{M.~C. Babiuc${}^{1}$, J. Winicour${}^{2,3}$
and Y. Zlochower${}^{4}$
       }
\affiliation{
${}^{1}$ Department of Physics \\
 Marshall University, Huntington, WV 25755, USA \\
${}^{2}$ Department of Physics and Astronomy \\
        University of Pittsburgh, Pittsburgh, PA 15260, USA\\
${}^{3}$ Max-Planck-Institut f\" ur
         Gravitationsphysik, Albert-Einstein-Institut, 
	 14476 Golm, Germany \\
${}^{4}$ Center for Computational Relativity and Gravitation and 
School of Mathematical Sciences\\
Rochester Institute of Technology, Rochester, NY 14623
}

\begin{abstract}

In this work, we present a work in progress towards an efficient and economical computational module which interfaces between Cauchy and characteristic evolution codes. Our goal is to provide a standardized waveform extraction tool for the numerical relativity community which will allow CCE to be readily applied to a generic Cauchy code. The tool provides a means of unambiguous comparison between the waveforms generated by evolution codes based upon different formulations of the Einstein equations and different numerical approximation.

\end{abstract}


\maketitle

\section{Introduction}

The study of gravitational waves will brighten  unexplored features of our universe that are otherwise invisible to conventional astronomy and will increase our knowledge about the very nature of time and space~\cite{shultz}. Gravitational wave detectors are already operating, and results from the first LIGO and Virgo collaboration were recently published in Nature~\cite{collab}. The signal predicted by numerical relativity will provide a template bank used for filtering the noise, indispensable to the success of gravitational wave detectors such as LIGO, Virgo, and LISA. The current sensitivity levels of the detectors will be improved substantially in next-generation detection estimated by 2015. Although existing simulations are sufficiently accurate for populating the parameter space in current searches of ground-based detectors, the new generation of advanced detectors will be 10-15 times more sensitive by 2015.

Ideally, the emitted gravitational wave signature should be extracted at spatial or null infinity. However, most present codes impose artificial, finite outer boundaries and are performing the waveform extraction at finite radius. This method introduces systematic errors, especially for higher modes, which is the main obstacle in reaching the desired accuracy. With the exception of Pretorius~\cite{pretorius}, who uses coordinates that compactify spatial infinity, all the other codes use a computational domain with a finite outer boundary and Sommerfeld-like approximate outer boundary conditions must be imposed, which introduce errors in the computation of gravitational waves. The choice of proper boundary conditions is complicated by gauge freedom and constraint preservation~\cite{lehnmor}. 
The emitted gravitational wave signature is calculated at finite distance, using either the Newman-Penrose Weyl scalar $\psi_4$~\cite{NP}, or the odd and even parity functions $Q_+$, $Q_x$ in the Zerilli-Moncrief formalism~\cite{moncrief}. The strain $h$ of the wave used in detection is obtained performing one time integration from the Zerilli-Moncrief multipoles, and two time integrations from the Newman-Penrose curvature. The waveform is affected by gauge ambiguities which are magnified by the integration~\cite{lindblom}.

Cauchy-characteristic extraction (CCE)~\cite{cce}, which is one of the pieces of the CCM strategy~\cite{livccm}, offers a means to avoid the error introduced by extraction at a finite world-tube. In CCE, the inner world-tube data supplied by the Cauchy evolution is used as boundary data for a characteristic evolution to future null infinity $\cal I^+$, where the waveform can be unambiguously computed by geometric methods.
This characteristic initial-boundary value problem based upon a timelike world-tube~\cite{tam} has been implemented as a mature evolution code, the PITT null code~\cite{isaac,highp}, which incorporates a Penrose compactification of the space-time. By itself, CCE does not use the characteristic evolution to inject outer boundary data for the Cauchy evolution, which can be a source of instability in full CCM. 

The PITT code has been tested  to be second order convergent in a wide range of testbeds extending from the perturbative regime~\cite{babiuc05} to highly
nonlinear single black hole spacetimes~\cite{highp}. However, in cases which
require high resolution, such as the inspiral of matter into a black
hole, the error in CCE has been a troublesome factor in the
postprocessing phase~\cite{partbh}. This has motivated a recent project~\cite{strat} to increase the accuracy of the PITT code. Other results achieved with previous versions of the PITT have been recently reported~\cite{reis1,reis2}. Recently, the code underwent major improvements and corrections to previous versions, to improve accuracy and convergence~\cite{xtract}.

Here we test this improved version of CCE on a realistic application involving a Cauchy evolution of the inspiral and merger of two equal mass non-spinning black holes. We use the same code specifications described in~\cite{strat} except that  the accuracy of angular derivatives has been increased to a 4th order finite difference approximation. The results presented here are a work in progress towards our goal to develop CCE as a reliable and accurate waveform extraction tool for the numerical relativity community.
This paper addresses the first two objectives: 
\begin{itemize}
\item To create a robust and flexible interface between a binary black-hole Cauchy evolution code and a characteristic code for wave extraction at infinity. 
\item To prove the robustness of the interface by performing precise computations of gravitational waveforms at infinity from binary black-hole, using this Cauchy-characteristic extraction approach.
\end {itemize}
We construct an interface that takes the Cartesian data from a Cauchy evolution and converts it into boundary data on a spherical grid for the characteristic evolution. The data are evolved to future null infinity, where it is used to compute the gravitational waveform.
The flexibility of the interface is due to implementation of a spectral decomposition of data. This implementation has been tested with a realistic application involving a binary black-hole inspiral. 
In Sec.~\ref{sec:formalism} we review the formalism underlying CCE,  including enough details of the patching, evolution and extraction, to make clear the difficulties underlying the calculation of an accurate waveform at ${\cal I}^+$.
In Sec.~\ref{sec:initdata} we briefly describe the initial data for Cauchy and characteristic evolution.
In Sec.~\ref{sec:interface}, we present the details of the CCE interface which allows the data from a Cauchy evolution to be used as boundary  data on an inner worltube for a characteristic evolution to ${\cal I}^+$, where the waveform is extracted.
In Sec.~\ref{sec:results}, we test the CCE interface by extracting the waveform from a Cauchy evolution of a binary black-hole inspiral and merger, and by comparing it to the waveform obtained by other standard method in current practice.

\section{The CCE Formalism}
\label{sec:formalism}

\subsection{Cauchy-Characteristic Patching}
\label{sec:cform}
Characteristic data are provided by the Cauchy evolution on a world-tube ${\cal WT}$, free initial data being given on the initial null hypersurface ${\cal N_I}$, which sets the metric on the entire initial cone (fig. 1).

 \begin{figure}[htp] 
   \centering
    \includegraphics*[width=6cm]{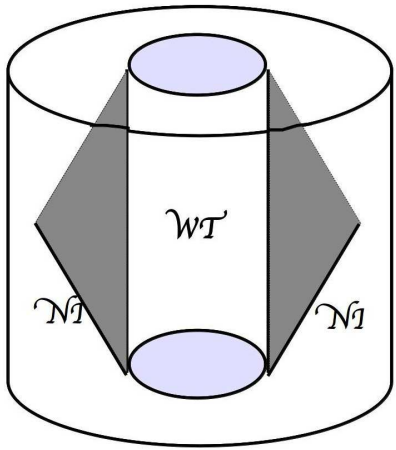}
    \caption{Cauchy and characteristic evolution are patched in the vicinity of a world-tube ${\cal WT}$, embedded in Cauchy evolution}
    \label{fig:Patching}
 \end{figure}    

The metric data from a Cauchy evolution are interpolated onto a timelike inner world-tube to extract the boundary data for the characteristic evolution. The characteristic evolution is embedded into the Cauchy evolution and is extending to future null infinity ${\cal I}^+$, where the waveform can be unambiguously computed using the geometric methods developed by Bondi et al~\cite{bondi}, Sachs~\cite{sachsr} and Penrose~\cite{Penrose}. The extraction process involves carrying out the complicated Jacobian transformation between the Cartesian coordinates used in the Cauchy evolution and the spherical null coordinates used in the characteristic evolution (the full details are given in~\cite{ccm}.)

\subsection{Characteristic Evolution}
\label{sec:cform}
The characteristic formalism is based upon a family of outgoing null
hypersurfaces, emanating from some inner world-tube, which extend to
infinity where they foliate ${\cal I}^+$ into spherical slices.
We let $u$ label these hypersurfaces, $x^A$ $(A=2,3)$ be angular coordinates
which label the null rays and $r$ be a surface area coordinate. (fig. 2).

 \begin{figure}[htp] 
   \centering
    \includegraphics*[width=6cm]{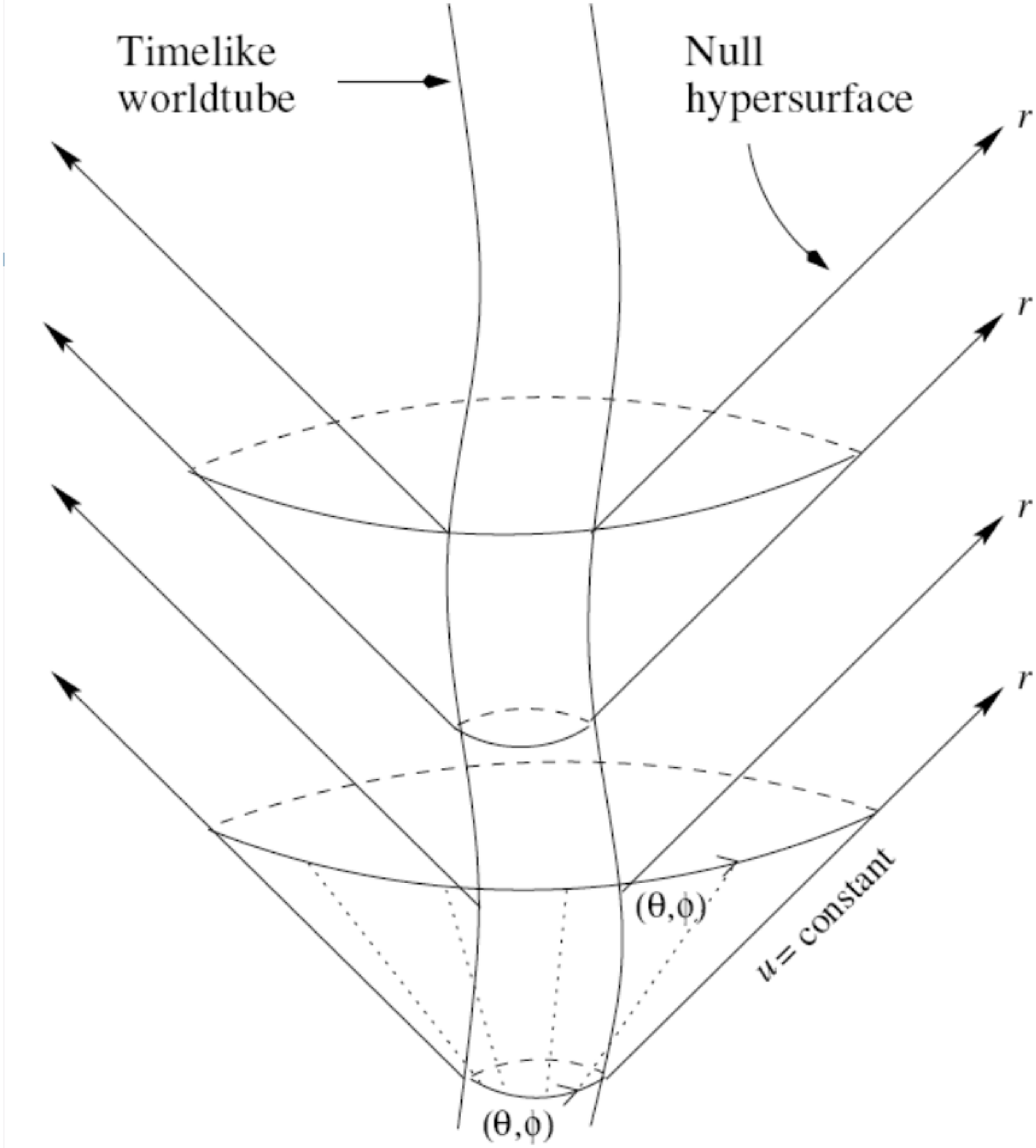}
    \caption{Ongoing null hypersurfaces emanating from the world-tube and extending to ${\cal I}^+$}
    \label{fig:Timelike}
 \end{figure}    

 In the resulting $x^\alpha=(u,r,x^A)$ coordinates, the metric takes the Bondi-Sachs form~\cite{bondi,sachsr}
\begin{eqnarray}
   ds^2 & = & -\left(e^{2\beta}\frac{V}{r} -r^2h_{AB}U^AU^B\right)du^2
        -2e^{2\beta}dudr \nonumber \\
       & -& 2r^2 h_{AB}U^Bdudx^A
         +  r^2h_{AB}dx^Adx^B,    
        \label{eq:bmet}
\end{eqnarray}
where  $h_{AB}$ is the Bondi-Sachs conformal 2-metric with $h^{AB}h_{BC}=\delta^A_C$. 
The code introduces an auxiliary unit sphere metric $q_{AB}$, with
associated complex dyad $q_A$ satisfying
$ q_{AB} =\frac{1}{2}\left(q_A \bar q_B+\bar q_Aq_B\right)$.
For a general Bondi-Sachs metric, the full nonlinear $h_{AB}$ is uniquely determined by the dyad component 
$J=h_{AB}q^Aq^B/2$, since the other dyad component $K=h_{AB}q^A \bar q^B /2$ is constrained by the determinant condition
$1=K^2-J\bar J$.  The spherically symmetric case characterized by $J=0$.
We introduce the spin-weighted fields $U=U^Aq_A$ and $Q=Q_Aq^A$, where 
\begin{equation}
     Q_A = r^2 e^{-2\,\beta} h_{AB} U^B_{,r}.
\end{equation}
as well as the (complex differential) operators $\eth$ and $\bar \eth$.
Refer to~\cite{eth,cce} for further details regarding numerical implementation. The auxiliary variables
\begin{equation}
     \nu =\eth J \, , \quad B=\eth \beta \, , \quad k=\eth K
\label{eq:aux}
\end{equation}
are also introduced to eliminate all second angular derivatives.
In certain applications this has been found to give rise to increased
accuracy by suppressing short wavelength error~\cite{gomezfo}. 

In this formalism, the Einstein equations $G_{\mu\nu}=0$ decompose into
hypersurface equations, evolution equations and conservation conditions on the
inner world-tube. As described in more detail in~\cite{newt,nullinf}, the
hypersurface equations take the form
\begin{eqnarray}
      \beta_{,r} &=& N_\beta, 
   \label{eq:beta} \\
          U_{,r}  &=& r^{-2}e^{2\beta}Q +N_U, 
     \label{eq:wua} \\
     (r^2 Q)_{,r}  &=& -r^2 (\bar \eth J + \eth K)_{,r}
                +2r^4\eth \left(r^{-2}\beta\right)_{,r} + N_Q, 
     \label{eq:wq} \\
V_{,r} &=& \frac{1}{2} e^{2\beta}{\cal R} 
- e^{\beta} \eth \bar \eth e^{\beta}
+ \frac{1}{4} r^{-2} \left(r^4
                           \left(\eth \bar U +\bar \eth U \right)
                     \right)_{,r} + N_V,
 \label{eq:ww}
\end{eqnarray}
where~\cite{eth}
\begin{equation}
{\cal R} =2 K - \eth \bar \eth K + \frac{1}{2}(\bar \eth^2 J + \eth^2 \bar J)
          +\frac{1}{4K}(\bar \eth \bar J \eth J - \bar \eth J \eth \bar J)
     \label{eq:calR}
\end{equation}
is the curvature scalar of the 2-metric $h_{AB}$. Those equations have a hierarchical structure in $[J,\beta,Q,U,V]$ 
such that the right hand sides, e..g. $N_\beta[J]$ only depend upon previous variables and their derivatives intrinsic to the hypersurface. 

The evolution equation takes the form
\begin{eqnarray}
    2 \left(rJ\right)_{,ur}
    - \left(r^{-1}V\left(rJ\right)_{,r}\right)_{,r} = 
    -r^{-1} \left(r^2\eth U\right)_{,r}
    + 2 r^{-1} e^{\beta} \eth^2 e^{\beta}- \left(r^{-1} V \right)_{,r} J
    + N_J,
    \label{eq:wev}
\end{eqnarray}
where, $N_\beta$, $N_U$, $N_Q$, $N_V$ and $N_J$ are nonlinear terms  which
vanish for spherical symmetry. Expressions for these terms as complex
spin-weighted fields and a discussion of the conservation conditions are given
in~\cite{cce}.

The characteristic Einstein equations are evolved in a domain between an inner radial boundary at the interior world-tube, and an outer boundary at future null infinity.
The characteristic evolution code implements this formalism as an explicit
finite difference scheme, based upon the compactified radial coordinate
\begin{equation}
      \xi=\frac{r}{R_E +r}
\label{eq:compx}
\end{equation}
so that ${\xi}=1$ at ${\cal I}^+$. Here $R_E$ is a parameter based upon 
the extraction world-tube, which in the CCE module is chosen as the
radius of the extraction world-tube, as determined by $R^2=\delta_{ij}x^i
x^j$ in terms  of the Cartesian coordiates $x^i$ used in the Cauchy
evolution code. 
The boundary data for $J$, $\beta$, $U$, $Q$, and $V$ on the world-tube supply the integration constants for a radial numerical integration of the hypersurface Einstein equations.
The finite difference scheme for integrating the hypersurface and evolution
equations is based on the marching equation for a spherically symmetric scalar field $\Phi$:
\begin{equation}
   \Phi_{\bf N}-\Phi_{\bf W}-\Phi_{\bf E}+\Phi_{\bf S} = -{1\over 2}\int_{\Sigma} \left ({V \over r} \right)_{,r} {\Phi \over r} du dr
  \label{eq:march}
\end{equation} 
where the point N is the "new" point in the evolution scheme, and $V$ is defined by the spherically symmetric version of the Bondi-Sachs metric given above. The evolution scheme in the full gravitational case used to determine the metric at the next point on the null hypersurfaces is modeled after this example (see ~\cite{highp,gomezfo} for details).

 \begin{figure}[htp] 
   \centering
    \includegraphics*[width=6cm]{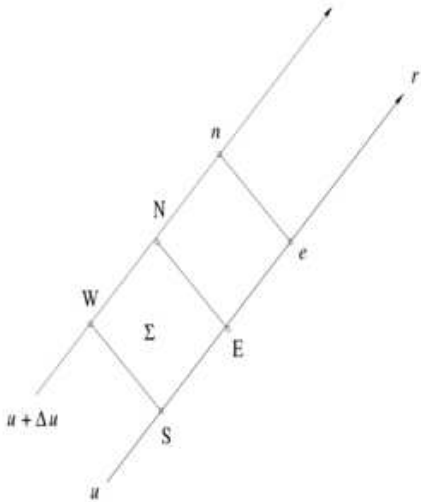}
    \caption{The null parallelogram WSEN used to determine the field values at point N, as  described by (\ref{eq:march}).}
    \label{fig:Null}
 \end{figure}

\subsection{Gravitational Radiation Calculation}
\label{sec:wave}

The theoretical derivation of the waveform at infinity is carried out in terms of an inverse surface-area coordinate 
$\ell=1/r$, where  $\ell=0$ at ${\cal I}^+$. In the resulting $x^\mu=(u,\ell,x^A)$ coordinates, the physical space-time 
metric $g_{\mu\nu}$ (\ref{eq:bmet}) has the conformal compactification $\hat g_{\mu\nu}=\ell^{2} g_{\mu\nu}$, 
where $\hat g_{\mu\nu}$ is smooth at ${\cal I}^+$ and takes the form~\cite{tam}
\begin{equation}
   \hat g_{\mu\nu}dx^\mu dx^\nu= 
           -\left(e^{2\beta}V \ell^3 -h_{AB}U^AU^B\right)du^2
        +2e^{2\beta}dud\ell -2 h_{AB}U^Bdudx^A + h_{AB}dx^Adx^B.
   \label{eq:lmet}
\end{equation}
As described in~\cite{strat}, the Bondi news function $N(u,x^A)$ and the Newman-Penrose Weyl tensor component
$\Psi(u,x^A)=\lim_{r\rightarrow \infty} r \psi_4$ 
which describe the waveform are both determined by the asymptotic limit at ${\cal I}^+$ of the tensor  field
\begin{equation}
 \hat \Sigma_{\mu\nu} = \frac{1}{\ell}(\hat \nabla_\mu\hat \nabla_\nu    -\frac{1}{4}\hat g_{\mu\nu} \hat \nabla^\alpha\hat \nabla_\alpha)\ell.
\label{eq:Sigma}
\end{equation}
constructed from the leading coefficients in an expansion of the metric in powers of $\ell$
\begin{eqnarray}
   h_{AB}&= &H_{AB}+\ell c_{AB}+O(\ell^2),
      \\
     \beta&=&H+ O(\ell^2) ,
     \\
     U^A&=& L^A+2\ell e^{2H} H^{AB}D_B H+O(\ell^2) ,
     \\
      \ell^2 V&=& D_A L^A+\ell (e^{2H}{\cal R}/2 +D_A D^A e^{2H})+O(\ell^2),
\end{eqnarray}
where ${\cal R}$ and $D_A$ are the 2-dimensional curvature scalar and
covariant derivative associated with $H_{AB}$. 

The expansion coefficients $H$, $H_{AB}$, $c_{AB}$ and $L^A$ (all
functions of $u$ and $x^A$) completely determine the radiation field.
Before the gravitational radiation is calculated from the metric in the
neighborhood of ${\cal I}^+$, it is necessary to determine the conformal factor $\omega$ relating  $H_{AB}$ to a unit sphere metric $Q_{AB}$, i.e.  to an inertial conformal Bondi frame~\cite{tam} satisfying
\begin{equation}
         Q_{AB}=\omega^2H_{AB}.
\label{eq:unsph}
\end{equation}
The news function $N(u,x^A)$ is directly computed by the code in terms
of the computational coordinates $(u,x^A)$, as opposed to the inertial
coordinates $(\tilde u,y^A)$ on ${\cal I}^+$ corresponding to an
idealized distant observatory. The transformation to inertial
coordinates proceeds first by introducing the conformally rescaled
metric $\tilde g_{\mu\nu} = \omega^2 \hat g_{\mu\nu}$ in which the
cross-sections of ${\cal I}^+$ have unit sphere geometry, in accord
with (\ref{eq:unsph}). Then the rescaled null vector $\tilde n^\mu =
\omega^{-1} \hat n^\mu$ is the generator of time translations on ${\cal
I}^+$, i.e.  $\tilde n^\mu \partial_\mu = \partial_{\tilde u}$. The
inertial coordinates thus satisfy the propagation equations
\begin{equation} 
      \hat n^\mu \partial_\mu \tilde u = \omega \, , \quad
        \hat n^\mu \partial_\mu y^A =0,
\label{eq:inertialc}
\end{equation} 
where $\hat n^\mu\partial_\mu =e^{-2H}(\partial_u + L^A
\partial_{x^A})$ in terms of the computational coordinates. The
inertial coordinates are obtained by integrating (\ref{eq:inertialc}),
thus establishing a second pair of stereographic grid patches
corresponding to $y^A$. Then the news function is transformed into
$N(\tilde u, y^A)$. 

The Bondi news function $N$ is given by (\ref{eq:news}),
\begin{equation}
    N={1\over 4}e^{-2i \delta}\omega^{-2}e^{-2H}F^A F^B
       \{(\partial_u+{\pounds_L})c_{AB}-{1\over 2}c_{AB} D_C L^C
        +2\omega D_A[\omega^{-2}D_B(\omega e^{2H})]\},
     \label{eq:news}
\end{equation}
where $\pounds_L$ is the Lie derivative with respect to $L^A$.
The Newman-Penrose Weyl tensor component $\Psi$ is given by (\ref{eq:psia})
\begin{equation}
      \Psi=\frac{1}{2} \omega^{-3}e^{-2i\delta}
    \hat n^\mu F^A F^B \bigg( 
        \partial_\mu  \hat \Sigma_{AB}
       -\partial_A \hat \Sigma_{\mu B}
       - \hat \Gamma^\alpha_{\mu B}\hat \Sigma_{A \alpha}
       +\hat \Gamma^\alpha_{A B}\hat \Sigma_{\mu\alpha}
                      \bigg)|_{\cal I^+} .
\label{eq:psia}
\end{equation}
In the inertial Bondi coordinates, the expression for the news function  (\ref{eq:news}) reduces to the simple form
\begin{equation}
    N={1\over 4}{\cal Q}^A {\cal Q}^B \partial_u c_{AB},
     \label{eq:inews}
\end{equation}
and (\ref{eq:psia}) reduces to the single term
\begin{equation}
         \Psi = \frac {1}{4} Q^A Q^B\partial_u^2  c_{AB} = \partial_u^2
               \partial_l J|_{{\cal I}^+} . 
\end{equation}
This is related to the expression for the news function in
inertial Bondi coordinates by
\begin{equation}
       \Psi =\partial_u N.
\label{eq:PsiNu}
\end{equation}
Equation (\ref{eq:PsiNu}) holds true in the linearized approximation of the Einstein equations. 
In the nonlinear case, the full expression for news and $\Psi$ must be used in the code.
This introduces additional challenges to numerical accuracy due to high order angular derivatives of
$\omega$ and large number of terms.

\section{Initial Data}
\label{sec:initdata}

\subsection{Initial Cauchy Data}
\label{sec:initCauchy}
For the Cauchy evolution we used the
LazEv code~\cite{Campanelli:2005dd, Zlochower:2005bj} along with the Cactus
framework~\cite{cactus_web} and Carpet~\cite{Schnetter:2003rb} mesh refinement
driver. LazEV is an eighth-order-accurate finite-difference code based upon the
Baumgarte-Shapiro-Shibata-Nakamura (BSSN) formulation~\cite{bssn1,bssn2} of Einstein's equations, which deals with the internal singularities by the moving puncture approach~\cite{Campanelli:2005dd, Baker:2005vv}. Our simulation used 9 levels of refinement with finest
resolution of $h= M/53.76$, and outer Cauchy boundary at $400M$. The initial
data consisted of a close quasicircular black-hole binary with orbital frequency
$M\Omega = 0.050$, leading to more than a complete orbit before merger
(see~\cite{Campanelli:2006uy}). We output the metric data on the extraction
world-tube every $\Delta t = M/32$.

\subsection{Initial Characteristic data}
\label{sec:initchar}

The initial data for the characteristic evolution consist of the values
of $J$ on the initial hypersurface $u=0$. One way of supressing incoming
radiation in the data would be to set the Newman-Penrose Weyl
tensor component $\Psi_0=0$ on the initial null hypersurface. For a
perturbation of the Schwarzschild metric, this condition implies no
incoming radiation in the linearized approximation. However, in order
to avoid shocks arising from incompatibility with the Cauchy data on
the extraction world tube $\xi=\xi_E$ (with $\xi$ given by \ref{eq:compx}), we also need to require that $J$ and
$\partial_\xi J$ are continuous. In the linearized approximation, the
condition that $\Psi_0=0$ implies that $\partial^2_\xi J=0$. The combination of those requirements 
leads to $J=J|_{\xi_E} + (\partial_\xi J)|_{\xi_E} (\xi-\xi_E)$, which would
imply that $J\ne 0$ at ${\cal I}^+$. For technical simplicity we avoid this
complication by initializing $J$ according to 
\begin{equation}
    J = J|_{\xi_E}\frac {(\xi-1)}{(\xi_E-1)},
    \label{eq:initJ}
\end{equation} 
which matches the Cauchy data and the derivatives at $\xi=\xi_E$ and is
consistent with asymptotic flatness. Since this choice of $J$
vanishes at infinity, the initial slice of ${\cal I}^+$ has a unit
sphere metric so that the conformal factor has the simple
initialization $\omega(0,p,q) =1$.

Given the initial data (\ref{eq:initJ}), this leads to complete knowledge of the metric on the initial null cone. 
Then (\ref{eq:wev}) gives an expression for $J_{,ur}$, which is used to determine $J$ on the ``next'' null cone, 
so that the process can be repeated to yield the complete metric throughout the domain, which extends to ${\cal I}^+$.

\section{Computational interface}
\label{sec:interface}

We have designed an interface that takes Cartesian grid data from a
Cauchy evolution and converts it into boundary data for characteristic
evolution on a spherical grid extending to ${\cal I}^+$. We treat each
component $g_{\mu\nu}(t,x^i)$ of the Cauchy metric as a scalar function
in the $x^i$ Cartesian coordinates which are used in the $3+1$
evolution. 

In order to make the interface as flexible as possible for
future development as a  community tool for waveform extraction, we
have based it upon a spectral decomposition of the Cauchy data in the
region between two world tubes or radii $R=R_1$ and $R=R_2$, where
$R=\sqrt{\delta_{ij}x^i x^j}$ is the Cartesian coordinate radius. Then
at a given time $t=T$, we decompose $g_{\mu\nu}(T,x^i)$ in terms of
Tchebychev polynomials of the second kind $U_k(R)$ and spherical
harmonics $Y_{l m}(\theta,\phi)$, where $(\theta,\phi)$ are related
to $x^i/R$ in the standard way. The Tchebychev polynomials are
conventionally defined as functions $U_k(\tau)$ on the interval $-1 \le
\tau \le 1$. Here we map them to the interval $R_1 \le R \le R_2$ by
the transformation
$$
      \tau (R)=\frac{2R-R_1 -R_2}{R_2 -R_1} .
$$
where the extraction shell thickness is determined by the number $k_{Max}$ of
Tchebychev polynomials used. (In tests of binary black holes with mass 
M we use a relatively small range $R_2 -R_1=10M$, a larger value of  
$k_{Max}$ would be needed if the range were expanded).  
Thus, for $R_1<R<R_2$, we expand 
\begin{equation}
    g_{\mu\nu} (T,x^i) = \sum_{k l m} C_{\mu \nu [k l m]}
   U_k (R) Y_{l m}(\theta,\phi).
\end{equation}

For the applications to waveform extraction given in this paper, it is
sufficient to consider $l \le l_{Max}$, where $l_{Max} =6$, and
$k\le k_{Max}$, where $k_{Max} =6$. 
The coefficients $C_{\mu \nu [klm]}$ then allow us to reconstruct a
spherical harmonic decomposition of each component of the Cauchy 
metric on the extraction world-tube $R=R_E$, i.e. 
\begin{equation}
   g_{\mu\nu [l m]}(T,R_E)
         =  \sum_{k} C_{\mu \nu [k l m]} U_k (R_E) .
\end{equation}
This decomposition is carried out at a sequence of Cauchy time steps
$T_N=T_0 +N\Delta T$, where $\Delta T$ is chosen to be much smaller
than the characteristic time scale of the problem but, for purposes of
economy, larger than the time step used for the Cauchy evolution. 
A fifth-order polynomial interpolation is carried out locally over the
$T_N$ to provide characteristic boundary data at any time $t$ in
analytic form. 

The extraction module also requires the derivatives $\partial_t
g_{\mu\nu}$ and $\partial_R g_{\mu\nu}$ at the extraction world-tube.
The $t$-derivative is constructed by a fourth-order-accurate finite-difference 
stencil using the surrounding Cauchy times $t=T_N$. 
The $R$-derivative is obtained analytically, at each time
level $T_N$, by differentiation of the Tchebychev polynomials.

The spherical harmonic interpolator from the Cartesian to the spherical
coordinates is part of the extraction module, but its 
resolution is controlled by the Cauchy evolution. 

The stereographic coordinates $x^A=(q,p)$ used to label the outgoing null rays
in the Bondi metric are matched to the spherical coordinates $(\theta,\phi)$
induced by the Cartesian Cauchy coordinates on the extraction worldtube by a
standard transformation, using the conventions in~\cite{eth}. 
The value of the surface-area coordinate $r$ in the Bondi-Sachs metric
is obtained on the extraction world-tube from the 2-determinant of the
Cartesian metric on the surfaces $t=T_N,R=R_E$. As a result the radius of the 
Bondi coordinate $r\ne const$ on the extraction world-tube. 
The metric has to be calculated at a common value of the surface coordinate $r$, 
because the original Cauchy extraction was at constant R.
In order to make this calculation possible,  the transformation from Cartesian coordinates $(t, x^i)$
to Bondi-Sachs coordinates $(u,r,x^A)$ is carried out via an
intermediate Sachs coordinate system $(u,\lambda,x^a)$~\cite{sachsr}
where $\lambda$ is an affine parameter along the outgoing null rays.
The affine freedom allows us to set $\lambda=0$ on the extraction
world-tube $R=R_E$. After carrying out the Jacobian transformation from
$(t,x^i)$ to $(u,\lambda,x^A)$, the Cartesian metric and its first
derivatives at the extraction world-tube provide a first-order Taylor
expansion in $\lambda$ (about $\lambda =0$) of the null metric in Sachs
coordinates. The corresponding Taylor expansion of the metric in
Bondi-Sachs coordinates then follows from the computed value of $r$ and
$\partial_\lambda r$ at $\lambda=0$, which are obtained from the
2-determinant of the Cartesian metric.
In order to obtain a first-order Taylor expansion for the Bondi metric variable $\beta$, 
the hypersurface equation (\ref{eq:beta}) must be used to evaluate $\partial_r \beta$ 
at the extraction world-tube. All other metric variables are
then initialized consistent with second order accuracy. Taylor expansions are also 
needed to start up the radial integration equations for the auxiliary variables (\ref{eq:aux}) 
used to convert angular derivatives to first-order form. These expansions are obtained 
from applying the $\eth$-operator to the Taylor expansion of the underlying metric. 
This is a complicated process because the $\eth$ operator intrinsic to the $\lambda=0$
extraction world-tube is not the same as the $\eth$ operator intrinsic to the
$r=const$ Bondi spheres (see~\cite{xtract} for a discussion). The low order intermediate 
Taylor expansions limit the accuracy of the result.  A new approach that avoids 
entirely the use of the Taylor expansion and gives better accuracy is presented in ~\cite{xtract}. 
In the original approach, used for this paper, the resulting Taylor expansion of the 
evolution variables is used to fill the points of the Bond-Sachs grid to start the integration 
of the characteristic hypersurface and evolution equations (\ref{eq:beta}) - (\ref{eq:wev}). The
integration proceeds from the extraction world tube to ${\cal I}^+$ on a radial
grid based upon the compactified $x$-coordinate (\ref{eq:compx}). 

Domain of dependence considerations place a constraint between the
characteristic time step $\Delta u$ and the size of the characteristic
grid analogous to the CFL condition for the Cauchy evolution. 
For an estimate, consider the Minkowski space case with the conformally
rescaled metric
\begin{equation}
    ds^2=-\frac{(1-\xi)^2}{R_E^2} du^2 -\frac {2}{R_E} du d\xi
       +  q_{AB} dx^A dx^B
\end{equation}
where the unit sphere metric takes the form
\begin{equation}
    q_{AB} dx^A dx^B = \frac{4}{1+p^2 +q^2} (dp^2 + dq^2). 
\end{equation}
The past light cone is determined by 
\begin{equation}
    \frac {du}{R_E} = \frac {-d\xi 
       - \sqrt {d\xi^2+(1-\xi)^2 q_{AB} dx^A dx^B}} {(1-\xi)^2}.
\end{equation}
For typical characteristic grid parameters, $\Delta p = \Delta q =
\Delta \xi /4$, the resulting restriction is 
\begin{equation}
    \frac {|\Delta u|}{R_E} < 8 \Delta \xi 
\label{eq:cfl}
\end{equation}
For a Cauchy simulation of a binary black-hole 
system of total mass $M$ with timestep $\Delta t =M/32$ (sufficient to
describe the typical frequencies of a binary system), (\ref{eq:cfl})
leads to
\begin{equation}
    \frac {M}{256 R_E} <  \Delta \xi ,
\end{equation}
for the choice of characteristic timestep $\Delta u = \Delta t$. The
corresponding number of radial gridpoints must roughly satisfy $N_\xi < 128
R_E/M$. This places no limit of practical concern on the resolution of the
characteristic evolution even for the small extraction radius $R_E =20 M$. Thus,
for purposes of CCE, there are no demanding CFL restrictions.

The interface was debugged and calibrated using the analytic Schwarzschild metric 
in Kerr-Schild coordinates $(t,x^i)$,
\begin{equation}
         g_{\mu\nu} = \eta_{\mu\nu}+\frac{2m}{r}k_\mu k_\nu ,
\end{equation}
where $k_\mu=(-1,x^i/r)$. 

\section{Results}
\label{sec:results}

We present results for the characteristic extracted waveform either in terms
of $\Psi$, related to the Bondi news by $\Psi=\partial_u N$ in the linearized regime, 
or, when comparing to the perturbative waveform, in terms of the Newman-Penrose component
$\psi_4$. The relationship between the Cauchy and the characteristic waveforms is: $(R-2M)\psi_4=-2\bar \Psi$. We decompose the signal in $l=10$ spherical harmonic modes but, for illustrative purposes, we concentrate on the dominant $(2,2)$ and sub-dominant $(4,4)$ modes.
The Cauchy data were given at the extraction radii $R~=~20M,~50M,~100M$. The relationship between the Cauchy radius $R$ and the characteristic world-tube radius $R_E$ is: $R_E/R=1 + 1/R +1/(4R^2)$.
The  characteristic extraction module was run with the following specifications: angular gridpoints = radial gridpoints = $60,~120,~240$, and
timestep $\Delta u~=~8 \Delta t,~4 \Delta t,~2 \Delta t$, where $\Delta t=M/32$. The test was run until $t/M=385$, using $4^{th}$-order accurate angular derivatives, on stereographic patches  with circular boundaries and angular dissipation $\epsilon_{Jx} = 0.001$ (see \cite{strat} for details on how the angular dissipation is added to the evolution equation \ref{eq:wev}). 
The results are shown for the highest resolution.
Table~\ref{bbhconvwt} gives the convergence rates for the world-tube variables obtained with a small extraction radius $R_E=20M$ at a time corresponding to the peak of the signal ($t\approx 200M$).
The rates are given for the real and the imaginary part. All quantities are very close to second order convergent, including $J_{,x}$, which is the term which determines the waveform.
\begin{table}[htdp]
\caption{Convergence rates of the $l=2,m=2$ mode for the metric variables
measured near the peak of the signal ($t\approx200/M$) at the world-tube, for an extraction radius $R=20M$.}
\begin{center}
\begin{tabular}{|c|c|c|}
   \hline
   $Variable$ & $ Rate_{Re} $  & $ Rate_{Im} $ \\
   \hline
   $\beta$	&	$2.02$	&	$2.01$\\
   $J$	&	$2.03$	&	$2.00$\\
   $J_{,x}$	&	$2.04$	&	$1.99$\\
   $Q$	&	$2.02$	&	$2.04$\\
   $U$	&	$2.02$	&	$2.02$\\
   $W$	&	$2.01$	&	$2.04$\\
     \hline 
\end{tabular}
\end{center}
\label{bbhconvwt}
\end{table}%

We are reporting only first-order convergence rates at future null infinity $\cal I^+$ for the Bondi News $B$ and the Weyl complonent $\Psi$, but the error is relatively small ($0.5\%$ during the late inspiral). From an extraction point of view, these errors are smaller than the error in the Cauchy code data and are of little concern. The data are not convergent at early time when high frequencies dominate the error. For a thorough analysis of the causes for the first-order accurate results and major improvements to the code see~\cite{xtract}.

\begin{figure}[htp] 
    \centering
    \includegraphics*[width=12cm]{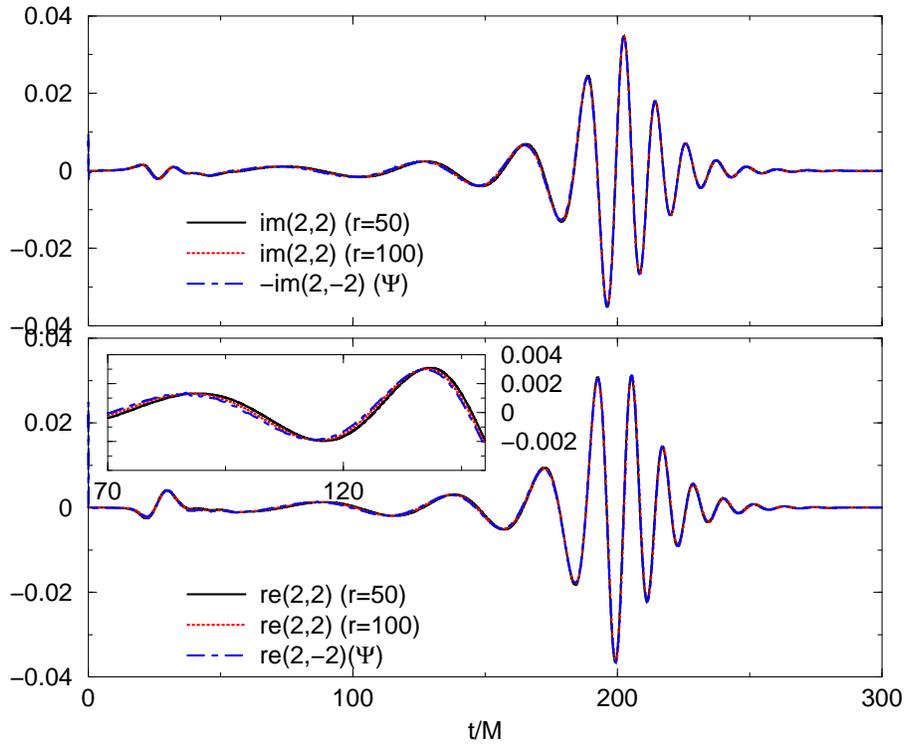}
    \caption{A plot that compares the phase in the $(l=2,m=2)$ mode
     of the Cauchy $\psi_4$ as calculated using the Null code and the
     Cauchy code. All plots were translated so that
     the time of the maximum in the amplitude agree.}
    \label{fig:Comp22}
 \end{figure}

Figure~\ref{fig:Comp22} compares the imaginary and real parts of the $(l, m) = (2, 2)$ mode of the Cauchy $\psi_4$
with the complex conjugate of the $(l, m) = (2,-2)$ mode of the characteristic $\Psi$. We obtain very good amplitude match. 
Also, we observe improved phase agreement as the extraction radius is increased (from $50M$ to $100M$), because the phase error in $\psi_4$ is reduced with increased extraction radius.

\begin{figure}[htp] 
    \centering
    \includegraphics*[width=12cm]{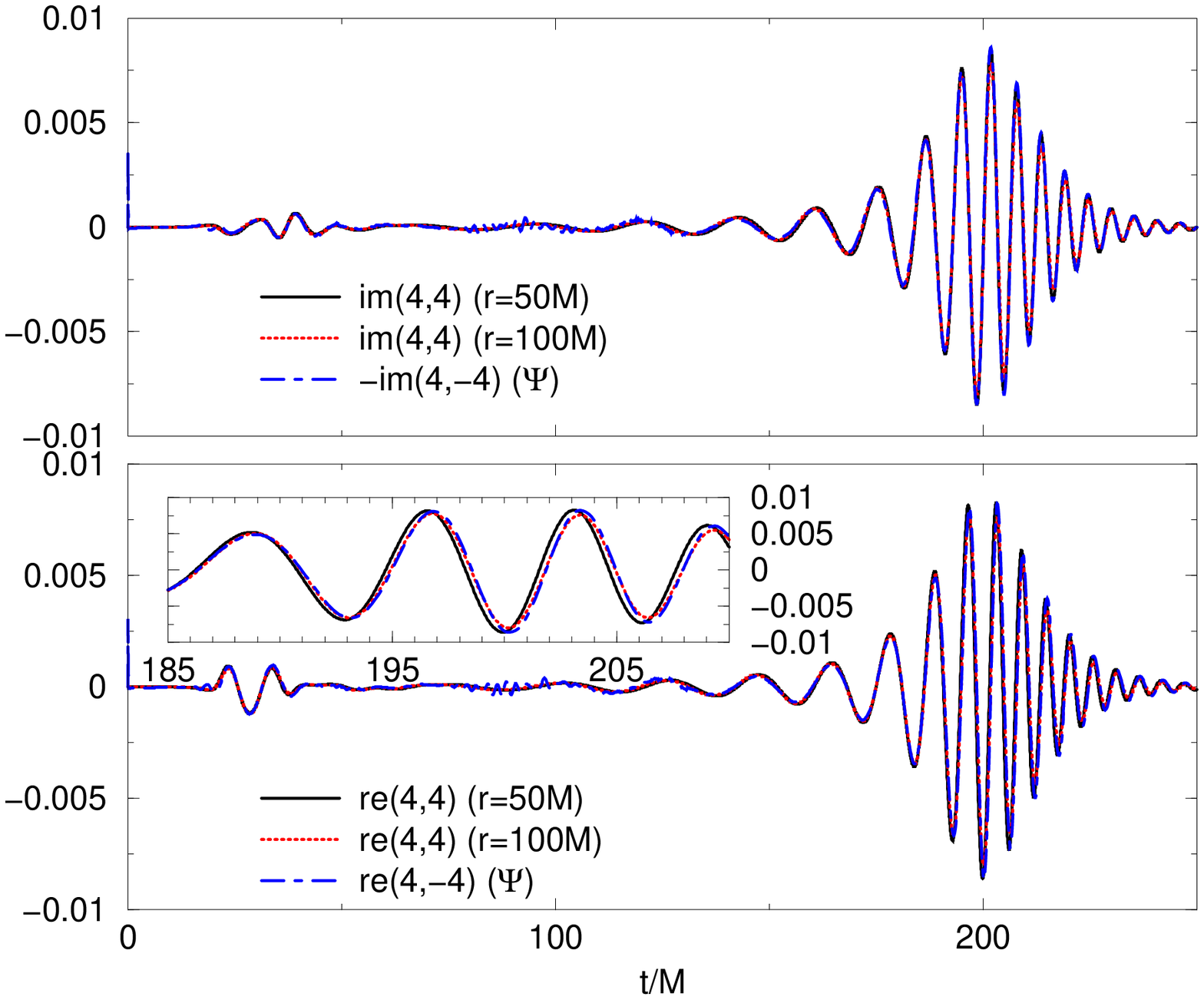}
    \caption{A plot that compares the phase in the $(l=4,m=4)$ mode
     of the Cauchy $\psi_4$ as calculated using the Null code and the
     Cauchy code. All plots were translated so that
     the time of the maximum in the amplitude agree.}
    \label{fig:Comp44}
 \end{figure}

Figure~\ref{fig:Comp44} compares the imaginary and real parts of the $(l, m) = (4, 4)$ mode of the Cauchy $\psi_4$ 
 with the complex conjugate of the $(l, m) = (4,-4)$ mode of the characteristic $\Psi$. Here we see two effects. First the improved phase agreement as $R_E\to \infty$, but also an attenuation of the amplitude due to
dissipation of higher-order modes. Also, the noise is apparent for the $(4,4)$ mode.

\begin{figure}[htp] 
    \centering
    \includegraphics*[width=12cm]{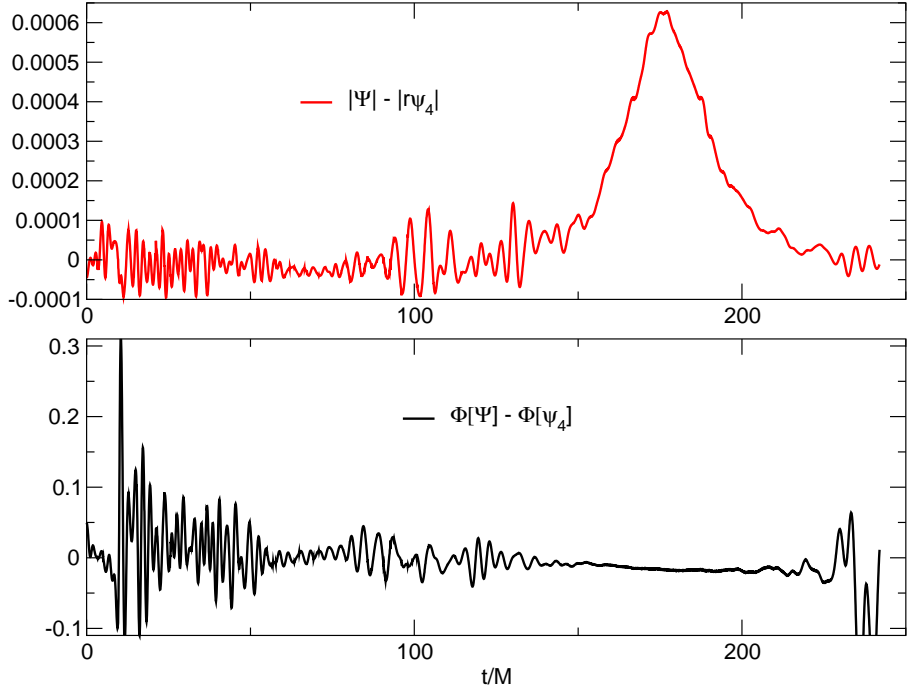}
    \caption{A plot of that compares the amplitudes and the phases between the $(l, m) = (2, 2)$ mode of the Cauchy $\psi_4$ extracted at $R=50$, and the $(l, m) = (2,-2)$ mode of the characteristic $\Psi$ for the same extraction radius.}
    \label{fig:CompAP}
 \end{figure}

Figure~\ref{fig:CompAP} compares the amplitudes and the phases between the 
absolute value of the $(l, m) = (2, 2)$ mode of the Cauchy $\psi_4$ extracted at $R=50$,
and the absolute value of the $(l, m) = (2,-2)$ mode of the characteristic $\Psi$ for the same extraction radius, 
at the highest resolution ($N=200$). The difference in amplitude is relatively small, 
maximum $0.17\%$  of the Cauchy $\psi_4$ amplitude in the wave zone. 

\begin{figure}[htp] 
    \centering
    \includegraphics*[width=12cm]{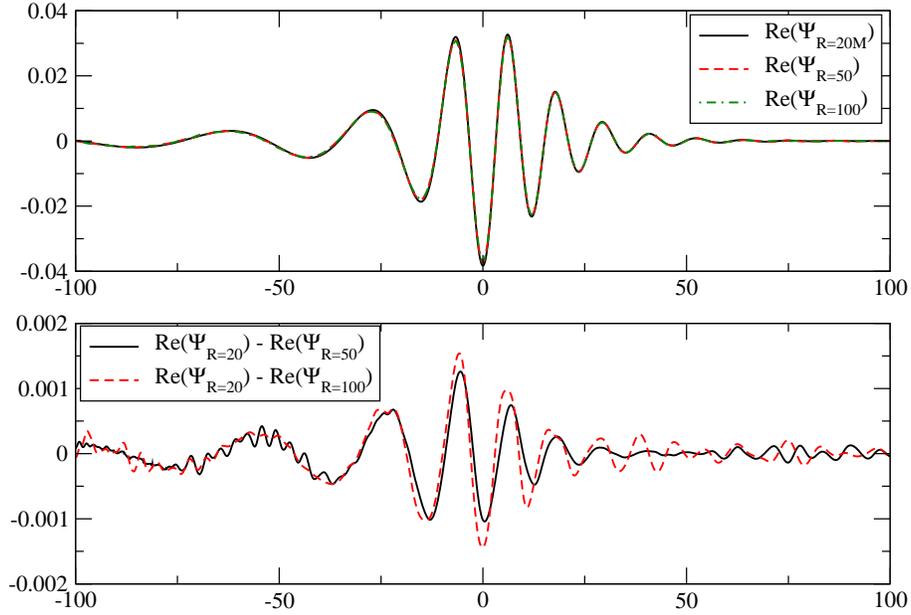}
    \caption{A plot of that compares the real part of the $(l, m) = (2,-2)$ mode for the characteristic $\Psi$ extracted at three different radius (world-tubes). The waveforms are translated such that the maximum of the amplitude corresponds to t/M=0}
    \label{fig:CompRad}
 \end{figure}

Figure~\ref{fig:CompRad} compares the real part of the $(l, m) = (2,-2)$ modes 
of the characteristic $\Psi$ extracted at three different extraction radii: 
$R=20M$, $R=50M$ and $R=100M$.
The waveform extracted at $R=20$ has the biggest amplitude, and a very small 
attenuation of the signal with the radius is observed. 

\section{Conclusion}
We have presented here a method for interfacing outer boundary data from a 
Cauchy evolution with inner boundary data for a characteristic evolution so that
the waveform can be accurately extracted at infinity. 
We have demonstrated how the PITT null code can be interfaced with the
LazEv code, which is a finite-difference BSSN code, to produced calibrated
waveforms from a binary black-hole inspiral. The extraction interface has been implemented 
as a thorn in the Einstein Computational Toolkit~\cite{einsteintool}
In this paper we are reporting only preliminary results (see~\cite{xtract} improvements).
 Although we are aware of the deficiencies in the characteristic waveform extraction
tool presented here, there is pressing interest from several numerical
relativity groups to apply the tool to extract waveforms from binary black-hole
inspirals.

\end {document}